\documentclass[reqno,10pt]{amsart}

\usepackage{amsmath}
\usepackage{amsfonts}   
\usepackage{amssymb}    
\usepackage{verbatim}
\usepackage{color, graphicx, graphics, pslatex, subfigure}

\theoremstyle{plain}
\newtheorem{lemma}{Lemma}
\newtheorem{theorem}[]{Theorem}
\newtheorem{corollary}[]{Corollary}

\theoremstyle{remark}

\newcommand*  {\N} {{\mathbb N}}

\newcommand*  {\R} {{\mathbb R}}

\newcommand*  {\C} {{\mathbb C}}

\newcommand*{\norm}[3][{\vphantom 1}]{\lVert #2 \rVert_{#3}^{#1}}

\newcommand*{\abs}[1]{\lvert #1 \rvert}

\newcommand{\nm}   [1]{\norm{#1}{}{}}
\newcommand{\ddt}     {\frac{d}{dt}}
\newcommand{\pdt}  [1]{\frac{d #1}{d t}}
\newcommand{\half}    {\frac{1}{2}}

\newcommand{\B}[1]{{\textbf{#1}}}

\begin{document}

\title[The dimension of the global attractor of the Sabra
shell model of turbulence]%
{Sharp lower bounds for the dimension of the global attractor of the
Sabra shell model of turbulence}

\date{October 31, 2006}



\author[P. Constantin]{Peter Constantin}
\address[P. Constantin]%
{Department of Mathematics \\
 The University of Chicago \\
 Chicago, IL 60637 \\
 USA}
\email{const@math.uchicago.edu}

\author[B. Levant]{Boris Levant}
\address[B. Levant]%
{Department of Computer Science and Applied Mathematics \\
 Weizmann Institute of Science \\
 Rehovot, 76100 \\
 Israel}
\email{boris.levant@weizmann.ac.il}

\author[E. S. Titi]{Edriss S. Titi}
\address[E. S. Titi]%
{Department of Mathematics and
 Department of Mechanical and Aerospace Engineering \\
 University of California \\
 Irvine, CA 92697 \\
 USA \\
 Also, Department of Computer Science and Applied Mathematics \\
 Weizmann Institute of Science \\
 Rehovot, 76100 \\
 Israel}
\email{etiti@math.uci.edu and edriss.titi@weizmann.ac.il}

\begin{abstract}
In this work we derive a lower bounds for the Hausdorff and fractal
dimensions of the global attractor of the Sabra shell model of
turbulence in different regimes of parameters. We show that for a
particular choice of the forcing and for sufficiently small
viscosity term $\nu$, the Sabra shell model has a global attractor
of  large Hausdorff and fractal dimensions proportional to
$\log_\lambda \nu^{-1}$ for all values of the governing parameter
$\epsilon$, except for $\epsilon=1$. The obtained lower bounds are
sharp, matching the upper bounds for the dimension of the global
attractor obtained in our previous work. Moreover, we show different
scenarios of the transition to chaos for different parameters regime
and for specific forcing. In the ``three-dimensional'' regime of
parameters this scenario changes when the parameter $\epsilon$
becomes sufficiently close to $0$ or to $1$. We also show that in
the ``two-dimensional'' regime of parameters for a certain non-zero
forcing term the long-time dynamics of the model becomes trivial for
any value of the viscosity.
\end{abstract}

\maketitle

\medskip


\section{Introduction}

Shell models of turbulence have attracted  interest as useful
phenomenological models that retain certain features of the
Navier-Stokes equations (NSE). In this work we continue our
analytical study of the Sabra shell model of turbulence that was
introduced in \cite{LP98}. For other shell models see \cite{BJPV98},
\cite{Gl73}, \cite{Fr95}, \cite{OY89}. A recent review of the
subject emphasizing the applications of the shell models to the
study of the energy-cascade mechanism in turbulence can be found in
\cite{Bi03}.

The Sabra shell model of turbulence describes the evolution of
complex Fourier-like components of a scalar velocity field denoted
by $u_n$. The associated one-dimensional wavenumbers are denoted by
$k_n$, where the discrete index $n$ is referred to as the ``shell
index''. The equations of motion of the Sabra shell model of
turbulence have the following form
\begin{equation} \label{eq_Sabra}
\frac{d u_n}{d t} = i (a k_{n+1} u_{n+2} u_{n+1}^* + b k_{n}
u_{n+1} u_{n-1}^* - c k_{n-1} u_{n-1} u_{n-2}) - \nu k_n^2 u_n +
f_n,
\end{equation}
for $n = 1, 2, 3, \dots$, with the boundary conditions $u_{-1} =
u_{0} = 0$. The wave numbers $k_n$ are taken to be
\begin{equation} \label{eq_freq}
k_n = k_0 \lambda^n,
\end{equation}
with $\lambda  > 1$ being the shell spacing parameter, and $k_0
> 0$. Although the equation does not capture any geometry, we will
consider $L = k_0^{-1}$ as a fixed typical length scale of the
model. In an analogy with the Navier-Stokes equations, $\nu > 0$
represents a kinematic viscosity and $f_n$ are the Fourier
components of the forcing.

The three parameters of the model $a, b$ and $c$ are real. In order
for the Sabra shell model to be a system of the hydrodynamic type we
require that in the inviscid ($\nu = 0$) and unforced ($f_n = 0$, $n
= 1, 2, 3, \dots$) case the model will have at least one quadratic
invariant. Requiring conservation of the energy
\begin{equation*} \label{eq_energy}
\mathrm{E} = \sum_{n=1}^\infty \abs{u_n}^2
\end{equation*}
leads to the following relation between the parameters of the model,
which we will refer to as an energy conservation condition
\begin{equation} \label{eq_energy_cons_assumption}
a + b + c = 0.
\end{equation}
Moreover, in the inviscid and unforced case the model possesses
another quadratic invariant
\begin{equation*} \label{eq_helicity}
\mathrm{W} = \sum_{n=1}^\infty \bigg( \frac{a}{c} \bigg)^n
\abs{u_n}^2.
\end{equation*}

The Sabra shell model (\ref{eq_Sabra}) has the following $6$
parameters: $\nu, \lambda, k_0, a, b$, and $c$. However, the
``characteristic length-scale'' $k_0^{-1}$ does not appear on its
own, but only in the following combinations: $k_0 a, k_0 b$, and
$k_0 c$. Therefore, without loss of generality we may assume that
$k_0 = 1$. Next, by rescaling the time
\begin{equation*}
t \to a t,
\end{equation*}
and using the energy conservation assumption
(\ref{eq_energy_cons_assumption}) we may set
\begin{equation*} \label{eq_Sabra_parameters}
a = 1, \;\;\; b = -\epsilon, \;\;\; c = \epsilon - 1.
\end{equation*}
Therefore, the Sabra shell model is in fact a three-parameter family
of equations with parameters $\nu > 0$, $\epsilon$, and $\lambda >
1$. In most of the numerical investigations of the shell models the
parameter $\lambda$ was set to $\lambda = 2$ (see \cite{BJPV98},
\cite{LP98}). The physically relevant range of parameters is
$\abs{a/c} > 1$, or equivalently, $0 < \epsilon < 2$ (see
\cite{LP98} for details). For $0 < \epsilon < 1$ the quantity
$\mathrm{W}$ is not sign-definite and therefore it is common to
associate it with the helicity -- in an analogy to the $3$D
turbulence. The $2$D parameters regime corresponds to $1 < \epsilon
< 2$, for which the quantity $\mathrm{W}$ becomes positive. In that
case the second conserved quadratic quantity $\mathrm{W}$ is
identified with the enstrophy -- in analogy to the $3$D turbulence.

Classical theories of turbulence assert that the turbulent flows
governed by the Navier-Stokes equations have finite number of
degrees of freedom (see, e.g., \cite{Fr95},
\cite{LL77_Hydrodynamics}). Arguing in the same vein one can state
that the sabra shell model with non-zero viscosity has finitely many
degrees of freedom. One of the ways to interpret such a physical
statement mathematically is to assume that the number of degrees of
freedom of the model corresponds to the Hausdorff or fractal
dimension of its global attractor. In our previous study of the
Sabra shell model of turbulence (\cite{CLT05}) we proved the
existence of a global attractor for the model and provided explicit
upper bounds of its Hausdorff and fractal dimensions. Therefore, we
proved that indeed the long-time dynamics of the sabra shell model
with non-zero viscosity has effectively finitely many degrees of
freedom. The question remains how many? The main motivation behind
this work is to provide a lower bound for the Hausdorff and fractal
dimensions of the global attractor. Namely, to show that for the
particular choice of the forcing term, and for all $\epsilon\in (0,
2)$, $\epsilon\neq 1$, the Hausdorff and fractal  dimensions of the
global attractor are large, proportional to the upper bound obtained
previously in \cite{CLT05}. However, we also give an example of the
forcing such that for $\epsilon\in (0, 1)$ and any non-zero
viscosity $\nu$, the long-time dynamics of the Sabra shell model of
turbulence is trivial.

In our work, we show that the Sabra shell model of turbulence
possesses a global attractor of large dimension for all values of
the parameter $\epsilon\in (0, 2)$, $\epsilon \neq 1$. In other
words we show that for every $\epsilon \neq 1$, the Hausdorff and
fractal dimensions of the attractor are proportional to
$\log_\lambda \nu^{-1}$ for small enough viscosity $\nu$. Therefore,
we extend and give a rigorous analytical justification for the
numerical results observed in \cite{YO87_3D} and \cite{YO88_2D} for
$\epsilon=1/2$, and $\epsilon=3/2$, which corresponds respectively
to the purely ``three'' and ``two-dimensional'' values of
parameters.

Moreover, in Section~\ref{sec:general}, we obtain an estimate of the
dimension of the global attractor in terms of the non-dimensional
generalized Grashoff number $G$ defined as
\begin{equation} \label{eq_grashof}
G = \frac{\abs{\B f}}{\nu^2 k_1^3},
\end{equation}
where $\abs{\B f}$ is an appropriate norm of the forcing term, which
will be defined later. More specifically, we show that for every
$\epsilon\in (0, 2)$, $\epsilon \neq 1$, and for a small enough
viscosity $\nu$, there exist positive constants $c_1, c_2, c_3$,
depending on $\lambda, \epsilon$, and independent of the viscosity
$\nu$ and the forcing term $\B f$, such that
\begin{equation} \label{eq:dimmmmm}
c_1 \log_\lambda G + c_2 \le \dim_H(\mathcal A) \le \dim_F(\mathcal
A) \le \half \log_\lambda G + c_3.
\end{equation}
The right hand side of the inequality was proved in \cite{CLT05},
and is true for every forcing term $\B f$. In this work we show that
this estimate is tight in the sense that for particular choices of
the forcing term $\B f$ the lower bound in (\ref{eq:dimmmmm}) is
achieved.

Furthermore, in Section~\ref{sec:single}, we study the linear
stability of the stationary solution of the Sabra shell model,
concentrated on a single mode $N$. We show that it becomes unstable
for every $N$ and for small enough viscosity for all $\epsilon\in
(0, 2)$, $\epsilon \neq 1$, thus correcting a result of
\cite{KoOkJe98}. By considering a stationary solution concentrated
on an infinite number of shells, we are able to demonstrate exactly
how the transition to chaos occurs both in the ``two'' and
``three-dimensional'' parameters regime, through successive
bifurcations, as the viscosity $\nu$ tends to zero.

In the ``three-dimensional'' regime $\epsilon\in (0, 1)$, when the
parameter $\epsilon$ becomes close to $0$ or $1$ the scenario of the
transition to chaos is different than in the rest of the interval.
Namely, for a fixed viscosity, when $\epsilon$ crosses the values
$0.05$ and $0.97$, the dimension of the unstable manifold of the
certain stationary solution drops by a factor of $3$. However, the
attractor in those regimes is still of the size proportional to
$\log_\lambda \nu^{-1}$, the chaotic behavior in the vicinity of the
particular stationary solution changes dramatically.

Finally in Section~\ref{sec:example}, we show that in the
``two-dimensional'' parameters regime the Sabra shell model has a
trivial attractor reduced to a single equilibrium solution for any
value of viscosity $\nu$, when the forcing is applied only to the
first shell. This result is similar to the one for the
$2$-dimensional NSE due to Yudovich \cite{Yu65_Example} and
independently by Marchioro \cite{Ma86_TrivialAttr}.

The transition to chaos in the GOY shell model of turbulence was
studied previously in \cite{BiLLP95} and \cite{KaLoSch97} by
investigating numerically the stability properties of the special
stationary solution corresponding to the single mode forcing, which
has a $k^{-1/3}$ Kolmogorov's scaling in the inertial range. It was
found that this solution becomes unstable at $\epsilon=0.3843$, and
at some value of $\epsilon$ the phase transition occurs, when many
stable directions become suddenly unstable. In this work we show
that the nature of the transition to chaos strongly depends on the
type of the forcing chosen.

First, we give a brief introduction to the mathematical formulations
of the Sabra shell model problem. More details on this subject could
be found in \cite{CLT05} and \cite{CLT05_Inviscid}.

\section{Preliminaries and Functional Setting}
\label{sec:preliminaries}

In this work we will consider the real form of the Sabra model
\begin{equation*} \label{sabrareal}
\frac{d u_n}{d t} = (a k_{n+1} u_{n+2} u_{n+1} + b k_{n} u_{n+1}
u_{n-1} + c k_{n-1} u_{n-1} u_{n-2}) - \nu k_n^2 u_n + f_n,
\end{equation*}
for $n = 1, 2, 3, \dots$, and $u_n$, $f_n$ are real for all $n$.
This formulation is obtained from the original one by assuming that
both the forcing $f_n$ and the velocity components $u_n$ in the
equation (\ref{eq_Sabra}) are purely imaginary. Our goal in this
work is to show that the upper bounds of the Hausdorff and fractal
dimensions of the global attractor of the Sabra shell model obtained
in \cite{CLT05} are optimal in the sense that they can be achieved
for some specific choice of the forcing. Therefore, this formulation
of the model is not restrictive, as long as we are able to show in
that case that the size of the global attractor matches the upper
bound of \cite{CLT05}.

Following the classical treatment of the NSE and Euler equations,
and in order to simplify the notation we write the system
(\ref{eq_Sabra}) in the following functional form
\begin{subequations} \label{eq_abstract_model}
\label{model}
\begin{gather}
\pdt{\B u} + \nu \B A \B u + \B B(\B u, \B u) = \B f \label{modeleq} \\
\B u(0) = \B u^{in}, \label{modelinit}
\end{gather}
\end{subequations}
in a Hilbert space $H$. The linear operator $\B A$ as well as the
bilinear operator $\B B$ will be defined below. In our case, the
space $H$ will be the sequences space $\ell^2$ over the field of
complex numbers $\R$. For every $\B u, \B v\in H$, the scalar
product $(\cdot, \cdot)$ and the corresponding norm $\abs{\cdot}$
defined as
\[
(\B u, \B v) = \sum_{n=1}^\infty u_n v_n, \;\;\; \abs{\B u} = \bigg(
\sum_{n=1}^\infty \abs{u_n}^2 \bigg)^{1/2}.
\]

The linear operator $\B A : D(\B A) \to H$ is a positive definite,
diagonal operator defined through its action on the sequences $\B u
= (u_1, u_2, \dots)$ by
\begin{equation*}
\B A \B u = (k_1^2 u_1, k_2^2 u_2, \dots),
\end{equation*}
were the eigenvalues $k_j^2$ satisfy the equation (\ref{eq_freq}).
Furthermore, we will need to define a space
\begin{equation*}
V := D(\B A^{1/2}) = \{ \B u = (u_1, u_2, u_3, \dots) \;:\;
\sum_{j=1}^\infty k_j^{2} \abs{u_j}^2 < \infty \}.
\end{equation*}

The bilinear operator $\B B(\B u, \B v) = (B_1(\B u, \B v), B_2(\B
u, \B v), \dots)$ will be defined formally in the following way. Let
$\B u = (u_1, u_2, \dots)$, $\B v = (v_1, v_2, \dots)$ be two
sequences, then
\begin{align*}
B_n(\B u, \B v) = - k_n \Big( \lambda v_{n+2} u_{n+1} - \epsilon
v_{n+1} u_{n-1} - \lambda^{-1} u_{n-1} v_{n-2} + \epsilon
\lambda^{-1} v_{n-1} u_{n-2} \Big),
\end{align*}
for $n=1, 2,\dots$, and where $u_0 = u_{-1} = v_0 = v_{-1} = 0$. It
is easy to see that our definition of $\B B(\B u, \B v)$ is
consistent with (\ref{eq_Sabra}). In~\cite{CLT05} we showed that
indeed our definition of $\B B(\B u, \B v)$ makes sense as an
element of $H$, whenever $\B u\in H$ and $\B v\in V$ or $\B u\in V$
and $\B v\in H$.

For more details on the material of this section see \cite{CLT05}
and \cite{CLT05_Inviscid}.

\section{Lower bounds for the dimension of the global attractor -- the ``two-dimensional'' parameter regime}
\label{sec:2d}

The Hausdorff and fractal dimensions of the global attractor of the
evolution equation are bounded from below by the dimension of the
unstable manifold of every stationary solution (see, e.g.,
\cite{BV83_AttractorDim}, \cite{Te88}). Therefore, in order to
derive the lower bound for the Hausdorff and fractal dimensions of
the global attractor of the Sabra shell model equation we will
construct a specific stationary solution of the equation
(\ref{eq_abstract_model}) and count the number of the linearly
unstable directions of that equilibrium. The same technique was
first used in \cite{MeSi61_Kolmogorov} (see also
\cite{BV83_AttractorDim}, \cite{Liu93_Attractor}) to obtain lower
bounds for the dimension of the Navier-Stokes global attractor in
$2$D. In this section we will consider the ``two-dimensional''
parameters regime of the Sabra shell model corresponding to
\begin{equation*} \label{2dregime}
1 < \epsilon < 2.
\end{equation*}
Consider the forcing
\begin{equation*} \label{eq:forcing}
\B f = (f_1, f_2, f_3, \dots),
\end{equation*}
where
\begin{equation} \label{eq:forcingdef}
f_n = \left\{
        \begin{array}{ll}
          k_n^\alpha, & n = 0 \mod 3, \\
          0, & \text{otherwise},
        \end{array}
      \right.
\end{equation}
for
\begin{equation} \label{alpha}
\alpha = \frac{1}{3} \log_\lambda \frac{\epsilon-1}{\epsilon} +
\frac{5}{3}.
\end{equation}
In order to avoid the questions of the existence and uniqueness of
the solutions to the problem (\ref{model}) (see \cite{CLT05} for
details) we will chose some large number $M
> 0$ such that $f_n = 0$, for all $n > M$. More precisely, such a
forcing is supported on the finite number of modes, and therefore,
according to the results of \cite{CLT05}, for every initial
conditions $\B u(0)\in H$ the unique solution of the Sabra shell
model of turbulence exists globally in time, and possess an
exponentially decaying dissipation range (see \cite{CLT05} for
details), in particular $u(t)\in V$ for all $t > 0$. It was also
established in \cite{CLT05} that for such a forcing $\B f$ the Sabra
shell model of turbulence has a global attractor, which is a compact
subspace of the space $V$. Later in this section we will specify how
large the number $M$ should be.

The corresponding stationary solution of the Sabra shell model
equation (\ref{eq_Sabra}) or (\ref{eq_abstract_model}) is
\begin{equation} \label{eq:equilibria}
\B u = (u_1, u_2, u_3, \dots),
\end{equation}
with
\begin{equation} \label{eq:equilibriadef}
u_n = \left\{
        \begin{array}{ll}
          \frac{f_n}{\nu k_n^2}, & n = 0 \mod 3, \\
          0, & \text{otherwise}.
        \end{array}
      \right.
\end{equation}

Consider $\B v = (v_1, v_2, v_3, \dots)\in H$ -- an arbitrary
perturbation of the stationary solution $\B u$. Plugging $\B u + \B
v$ into the equation of motion (\ref{eq_abstract_model}) we find
that the perturbation $\B v$ satisfies the equation
\begin{equation*} \label{eq:perturbation}
\pdt{\B v} + \nu \B A \B v + \B B(\B u, \B v) + \B B(\B v, \B u) +
\B B(\B v, \B v) = \B 0.
\end{equation*}
To study the linear stability of the equilibrium solution $\B u$ we
will consider the properties of the linearized equation
\begin{equation*} \label{eq:perturbationlinear}
\pdt{\B v} + \B L_{\B u} \B v = \B 0,
\end{equation*}
where the linear operator is defined as
\begin{equation} \label{eq:linear}
\B L_{\B u} \B v = \nu \B A \B v + \B B(\B u, \B v) + \B B(\B v, \B
u).
\end{equation}
We are looking for the solution of the eigenvalue problem
\begin{equation} \label{eq:eigenvalue}
\B L_{\B u} \B v = -\sigma \B v,
\end{equation}
for some $\sigma\in \C$. Our goal is to count the number of the
solution of equation (\ref{eq:eigenvalue}) with $Re(\sigma) > 0$.
The equation (\ref{eq:eigenvalue}) in the componentwise form can be
written as
\begin{align}
\nu k_n^2 v_n - k_n & \big( \lambda u_{n+2} v_{n+1} - \epsilon
u_{n+1} v_{n-1} + (\epsilon - 1) \lambda^{-1} u_{n-1} v_{n-2} +
\notag
\\
& + \lambda v_{n+2} u_{n+1} - \epsilon v_{n+1} u_{n-1} + (\epsilon -
1) \lambda^{-1} v_{n-1} u_{n-2} \big) = -\sigma v_n .
\label{eq:eigenvaluecomp}
\end{align}
where $u_n$ is specified in (\ref{eq:equilibriadef}). Note, that
$u_n = 0$ for all $n\neq 0 \mod 3$, therefore the last equation
could be written in the following detailed form
\begin{itemize}
\item For $n=0\mod 3$,
\begin{equation} \label{eq:0mod}
\nu k_n^2 v_{n} = -\sigma v_n.
\end{equation}

\item For $n=1\mod 3$,
\begin{equation} \label{eq:1mod}
\nu k_n^2 v_{n} - k_n \big( (\lambda u_{n+2} - \epsilon u_{n-1})
v_{n+1} + (\epsilon - 1) \lambda^{-1} u_{n-1} v_{n-2} \big) =
-\sigma v_n.
\end{equation}

\item For $n=2\mod 3$,
\begin{equation} \label{eq:2mod}
\nu k_n^2 v_{n} - k_n \big( \lambda u_{n+1} v_{n+2} + ((\epsilon -
1) \lambda^{-1} u_{n-2} - \epsilon u_{n+1}) v_{n-1} \big) = -\sigma
v_n.
\end{equation}
\end{itemize}

Note that from the relation (\ref{eq:0mod}) it follows that $\sigma
= -\nu k_{n_0}^2$ correspond to the eigenvectors
\begin{equation} \label{eq:0eig}
\B v = (0, \dots, 0, v_{n_0}, 0, \dots),
\end{equation}
with $v_{n_0}\neq 0$ for some $n_0 = 0\mod 3$. However, we are only
interested in the solutions of the equation (\ref{eq:eigenvalue})
with $Re(\sigma) > 0$. Based on the above the only solution of the
relation (\ref{eq:eigenvaluecomp}) for which $Re(\sigma) > 0$ should
satisfy $v_n = 0$, for all $n=0\mod 3$. The equations
(\ref{eq:0mod}) are not coupled with the rest of the recursive
equations (\ref{eq:1mod}) and (\ref{eq:2mod}). Therefore, in looking
for non-trivial solutions $\B v$ of the equation
(\ref{eq:eigenvalue}) we can look only for the coupled recursive
linear equations (\ref{eq:1mod}) and (\ref{eq:2mod}), and set
\begin{equation}
v_n = 0, \;\;\; \forall n = 0 \mod 3,
\end{equation}
as the solution of (\ref{eq:0mod}).

In what follows we will find sufficient conditions for the existence
of non-trivial solutions for (\ref{eq:eigenvalue}) with $Re(\sigma)
> 0$. Denote
\begin{align*}
b_{n, 1} & = k_n \frac{\lambda u_{n+2} - \epsilon u_{n-1}}{\nu k_n^2
+ \sigma}, \notag
\\
c_{n, 1} & = \frac{(\epsilon - 1) k_{n-1} u_{n-1}}{\nu k_n^2 +
\sigma},
\end{align*}
for all $n = 1\mod 3$, and
\begin{align*}
b_{n, 2} & = \frac{k_{n+1} u_{n+1}}{\nu k_n^2 + \sigma}, \notag
\\
c_{n, 2} & = k_n \frac{(\epsilon - 1) \lambda^{-1} u_{n-2} -
\epsilon u_{n+1}}{\nu k_n^2 + \sigma},
\end{align*}
for all $n = 2\mod 3$. Then we can rewrite equations (\ref{eq:1mod})
and (\ref{eq:2mod}) as a recursive relation for $v_n$
\begin{align}
v_n - b_{n,1} v_{n+1} - c_{n,1} v_{n-2} & = 0, \;\;\; \text{for all}
\;\; n = 1 \mod 3, \notag
\\
v_n - b_{n,2} v_{n+2} - c_{n,2} v_{n-1} & = 0, \;\;\; \text{for all}
\;\; n = 2 \mod 3. \notag
\end{align}
Due to our choice (\ref{alpha}) for the value of $\alpha$, one can
realize from (\ref{eq:equilibriadef}) that $c_{n, 2} = 0$, for all
$n=2\mod 3$. Therefore, we can further simplify the last equations,
which become
\begin{equation} \label{recursion}
\begin{split}
v_n - b_{n,1} v_{n+1} - c_{n,1} v_{n-2} & = 0, \;\;\; n = 1 \mod 3,
\\
v_n - b_{n,2} v_{n+2} & = 0, \;\;\; n = 2 \mod 3.
\end{split}
\end{equation}
The following result gives a sufficient condition for the last
recursion to have at least one non-trivial solution.

\begin{lemma} \label{lemma1}
Let $M$ be a large positive integer. Let us fix $N < M$, and assume
that $N = 1\mod 3$. Then the recursive equation (\ref{recursion})
has a non-trivial solution of the form $v_n = 0$, for all $n > N$,
and $v_n \neq 0$, for some $n\le N$, if and only if
\begin{equation} \label{condition}
b_{N-2,2}~ c_{N,1} = 1.
\end{equation}
\end{lemma}

\begin{proof}
The proof of the Lemma~\ref{lemma1} is simple once we observe that
the recursive relation (\ref{recursion}) for solutions of the form
$v_n = 0$, for $n > N$, becomes
\begin{align*}
& v_1 - b_{1,1} v_{2} = 0,
\\
& v_{2} - b_{2,2} v_{4} = 0,
\\
& v_{4} - b_{4,1} v_{5} - c_{4,1} v_{2} = 0,
\\
& v_{5} - b_{5,2} v_{7} = 0,
\\
\vdots
\\
& v_{N-2} - b_{N-2,2} v_{N} = 0,
\\
& v_{N} - c_{N,1} v_{N-2} = 0.
\end{align*}
The last two equations have a one-parameter family of nontrivial
solutions if and only if the condition (\ref{condition}) is
satisfied.
\end{proof}

Finally, we are ready to prove the main result of this section.

\begin{theorem}
The Hausdorff and fractal dimensions of the global attractor of the
equation (\ref{model}) in the parameter regime $1 < \epsilon < 2$,
with the forcing term $\B f$ specified in (\ref{eq:forcingdef}),
satisfy
\begin{equation} \label{lower2d}
\dim_F \mathcal{A} \ge \dim_H \mathcal{A} \ge \frac{2}{4 -
\log_\lambda \frac{\epsilon - 1}{\epsilon}} \log_\lambda \nu^{-1} +
\frac{1}{8 - 2 \log_\lambda \frac{\epsilon - 1}{\epsilon}}
\log_\lambda (\epsilon - 1).
\end{equation}
\end{theorem}

\begin{proof}
Fix $M$ to be large enough, and let $N < M$ be such that $N = 1\mod
3$. Suppose, that for such $N$ the condition (\ref{condition}) is
satisfied for certain $\sigma=\sigma(N)$, depending on $N$, for
which $Re(\sigma) > 0$. Then, for such $\sigma$, there exists a
solution of equation (\ref{recursion}), and in particular, there
exists a solution of the eigenvalue problem (\ref{eq:eigenvalue})
with $Re(\sigma) > 0$. Moreover, it is not hard to see that if
$N_1\neq N_2$, then the solutions of the eigenvalue problem
(\ref{eq:eigenvalue}) corresponding to $\sigma(N_1)$ and
$\sigma(N_2)$ are different.

Therefore, for a given $M$, in order to count the number of unstable
directions of the stationary solution (\ref{eq:equilibriadef}), we
need to count the number of $N$-s, such that (i) $N < M$; (ii) $N =
1\mod 3$; (iii) $N$ satisfies (\ref{condition}) with the eigenvalue
$\sigma$, for which $Re(\sigma) > 0$.

Let us fix $N > 0$, satisfying $N = 1\mod 3$. The condition
(\ref{condition}) becomes
\begin{equation*}
\frac{k_{N+1} u_{N+1}}{\nu k_N^2 + \sigma} \cdot \frac{(\epsilon -
1) k_{N+1} u_{N+1}}{\nu k_{N+2}^2 + \sigma} = 1.
\end{equation*}
We get the quadratic equation in $\sigma$
\begin{equation*}
\sigma^2 + (\nu k_{N}^2 + \nu k_{N+2}^2) \sigma + \nu^2 k_{N}^2
k_{N+2}^2 - (\epsilon - 1) k_{N+1}^2 u_{N+1}^2 = 0.
\end{equation*}
This equation has a real positive solution, provided
\begin{equation} \label{cond1}
\nu^2 k_{N}^2 k_{N+2}^2 - (\epsilon - 1) k_{N+1}^2 u_{N+1}^2 < 0.
\end{equation}
Substituting (\ref{eq:equilibriadef}) we obtain the equivalent
condition to (\ref{cond1})
\begin{equation*}
(\epsilon - 1) \nu^{-2} k_{N+1}^{2(\alpha - 2)} > \nu^2 k_{N+1}^2.
\end{equation*}
Rearranging terms, the following conditions guarantees the existence
of a positive real eigenvalue for (\ref{eq:eigenvalue})
\begin{equation*}
(\epsilon - 1)^{1/2} \nu^{-2} > \lambda^{(3 - \alpha)(N+1)}.
\end{equation*}
Now, we substitute the value of $\alpha$ from (\ref{alpha}) to
obtain
\begin{equation*}
(\epsilon - 1)^{1/2} \nu^{-2} > \lambda^{\frac{4 - \log_\lambda
\frac{\epsilon - 1}{\epsilon}}{3}(N+1)}.
\end{equation*}
Finally, we get the estimate
\begin{align}
N+1 & < \frac{3}{4 - \log_\lambda \frac{\epsilon - 1}{\epsilon}}
\log_\lambda \big((\epsilon - 1)^{1/2} \nu^{-2} \big) = \notag
\\
& = \frac{6}{4 - \log_\lambda \frac{\epsilon - 1}{\epsilon}}
\log_\lambda \nu^{-1} + \frac{3}{8 - 2 \log_\lambda \frac{\epsilon -
1}{\epsilon}} \log_\lambda (\epsilon - 1). \label{eq:bndddd}
\end{align}

Therefore, we showed that if the $M$ that we have chosen at the
beginning of the proof, is larger than the right hand-side the
relation (\ref{eq:bndddd}), then for such a choice of the forcing
term, the number of the unstable direction of the stationary
solution (\ref{eq:equilibriadef}) is bounded from below by
\begin{equation*}
\frac{2}{4 - \log_\lambda \frac{\epsilon - 1}{\epsilon}}
\log_\lambda \nu^{-1} + \frac{1}{8 - 2 \log_\lambda \frac{\epsilon -
1}{\epsilon}} \log_\lambda (\epsilon - 1),
\end{equation*}
and the statement of the theorem follows.
\end{proof}

In \cite{CLT05} we showed that the dimension of the global attractor
of the Sabra shell model of turbulence is proportional to the
$\log_\lambda \nu^{-1}$ for small enough viscosity $\nu$. Therefore,
our result proves that this bound is tight.

\section{Lower bounds for the dimension of the global attractor -- the ``three-dimensional'' parameters regime}
\label{sec:general}

The result obtained in the previous section did not give an answer
for the case
\begin{equation}
0 < \epsilon < 1,
\end{equation}
which is also known as the ``three-dimensional'' range of
parameters. Therefore, we will need to apply different strategy.
First, we will consider the linear stability of a stationary
solution, corresponding to the force acting on a single mode number
$N$, for some $N > 0$. We will show that for every choice of $N$,
and for every value of the parameter $\epsilon \in (0, 2]$,
$\epsilon \neq 1$, such a stationary solution becomes linearly
unstable for sufficiently small viscosity $\nu$. The stability of a
single-mode stationary solution was numerically studied previously
in \cite{KoOkJe98}, where it was stated that such a solution becomes
stable around $\epsilon = 1$. Our rigorous proof contradicts this
numerical observation.

Next, we will construct a special type of an equilibrium solution,
for which we will be able to count the number of unstable
directions. The draw-back of this method is that we are not able to
obtain the exact dependance of the bounds on the parameters of the
problem $\epsilon$ and $\lambda$, as we succeeded in the
``two-dimensional'' parameters case.

\subsection{On the linear stability of a ``single-mode'' flow}
\label{sec:single}

Let us fix $N \ge 1$ and consider the forcing acting on the single
mode $N$ of the form
\begin{equation} \label{eq_single_mode_forcing}
\B f^N = (0, \dots, 0, \nu k_{N}^{-1}, 0, \dots),
\end{equation}
where all the components of $\B f^N$, except the $N$-th, are zero.
Consider one particular choice of an equilibrium  solution
corresponding to the above forcing
\begin{equation} \label{eq_kolmogorov_flow}
\B u^N = (0, \dots, 0, k_{N}^{-3}, 0, \dots),
\end{equation}
which is the analog of the Kolmogorov flow for the Navier-Stokes
equations.

Linearizing the equation (\ref{eq_Sabra}) around the equilibrium
solution $\B u^N$ and writing the equation (\ref{eq:eigenvalue}) in
the component form we get the following set of equations. For every
$j\in \N$, satisfying $2 < \abs{j - N}$, or $j = N$, we have
\begin{equation} \label{eq:trivial}
\nu k_j^2 v_j = -\sigma v_j,
\end{equation}
accompanied with the four equations, coming from the nonlinear
interaction with $\B u^N$
\begin{align*}
\nu k_{N-2}^2 v_{N-2} - k_{N-1} k_{N}^{-3} v_{N-1} & = -\sigma
v_{N-2},
\\
\nu k_{N-1}^2 v_{N-1} - k_{N} k_{N}^{-3} v_{N+1} + \epsilon k_{N-1}
k_{N}^{-3} v_{N-2} & = -\sigma v_{N-1}
\\
\nu k_{N+1}^2 v_{N+1} + (1 - \epsilon) k_N k_{N}^{-3} v_{N-1} +
\epsilon k_{N+1} k_{N}^{-3} v_{N+2} & = -\sigma v_{N+1},
\\
\nu k_{N+2}^2 v_{N+2} + (1 - \epsilon) k_{N+1} k_{N}^{-3} v_{N+1} &
= -\sigma v_{N+2}.
\end{align*}
Therefore, the eigenvalues of the linear operator $\B L_{\B u^N}$
(see (\ref{eq:linear})) are $-\sigma = \nu k_j^2$, for $2 < \abs{j -
N}$, or for $j = N$, corresponding to the eigenvectors $\B v = (0,
0, \dots, 1, 0, \dots)$, with $1$ at $j$-th place. Clearly, those
eigenvalues are positive, corresponding to $Re(\sigma) < 0$,
therefore they do not contribute to the number of the linearly
unstable directions of the equilibria $\B u^N$.

Other eigenvalues of the linear operator $\B L_{\B u^N}$ are the
eigenvalues of the following matrix
\begin{align} \label{eq_matrix_j}
J_N = \begin{pmatrix}
  \nu k_{N-2}^2 & -k_{N}^{-2} \lambda^{-1} & 0 & 0 \\
  \epsilon k_{N}^{-2} \lambda^{-1} & \nu k_{N-1}^2 & -k_{N}^{-2} & 0 \\
  0 & (1 - \epsilon) k_N^{-2} & \nu k_{N+1}^2 & \epsilon k_{N}^{-2} \lambda \\
  0 & 0 & (1 - \epsilon) k_{N}^{-2} \lambda & \nu k_{N+2}^2
\end{pmatrix},
\end{align}
which will correspond to the eigenvectors $\B v = (v_1, v_2, v_3,
\dots)$ of the linear operator $\B L_{\B u^N}$ with the only
non-zero components $v_j$, $0 < \abs{j - N} < 2$.

Our goal is to find the condition on the parameters $N, \epsilon$,
and $\nu$, such that the matrix $J_N$ has eigenvalues with the
negative real part, which will correspond to $\sigma$ satisfying
$Re(\sigma) > 0$. Let us rewrite the expression (\ref{eq_matrix_j})
it in the following way
\begin{align} \label{eq_matrix_j1}
J_N = k_N^{-2} \cdot \begin{pmatrix}
  \lambda^{-4} \beta & -\lambda^{-1} & 0 & 0 \\
  \epsilon \lambda^{-1} & \lambda^{-2} \beta & - 1 & 0 \\
  0 & (1 - \epsilon) & \lambda^{2} \beta & \epsilon \lambda \\
  0 & 0 & (1 - \epsilon) \lambda & \lambda^{4} \beta
\end{pmatrix},
\end{align}
where we denoted for simplicity
\[
\beta = \nu k_N^{4}.
\]
First, by substituting $\epsilon = 1$, we find that for this value
of $\epsilon$ the eigenvalues of the matrix $J_N$ has always
positive real part. Therefore, we conclude that in the case of
$\epsilon = 1$ the solution $\B u^N$ is stable for every $N$ and any
$\nu$.

\begin{figure}[!ht] \label{figure1}
\begin{center}
\subfigure[``Three-dimensional'' parameter regime.] {
\includegraphics[scale=0.5]{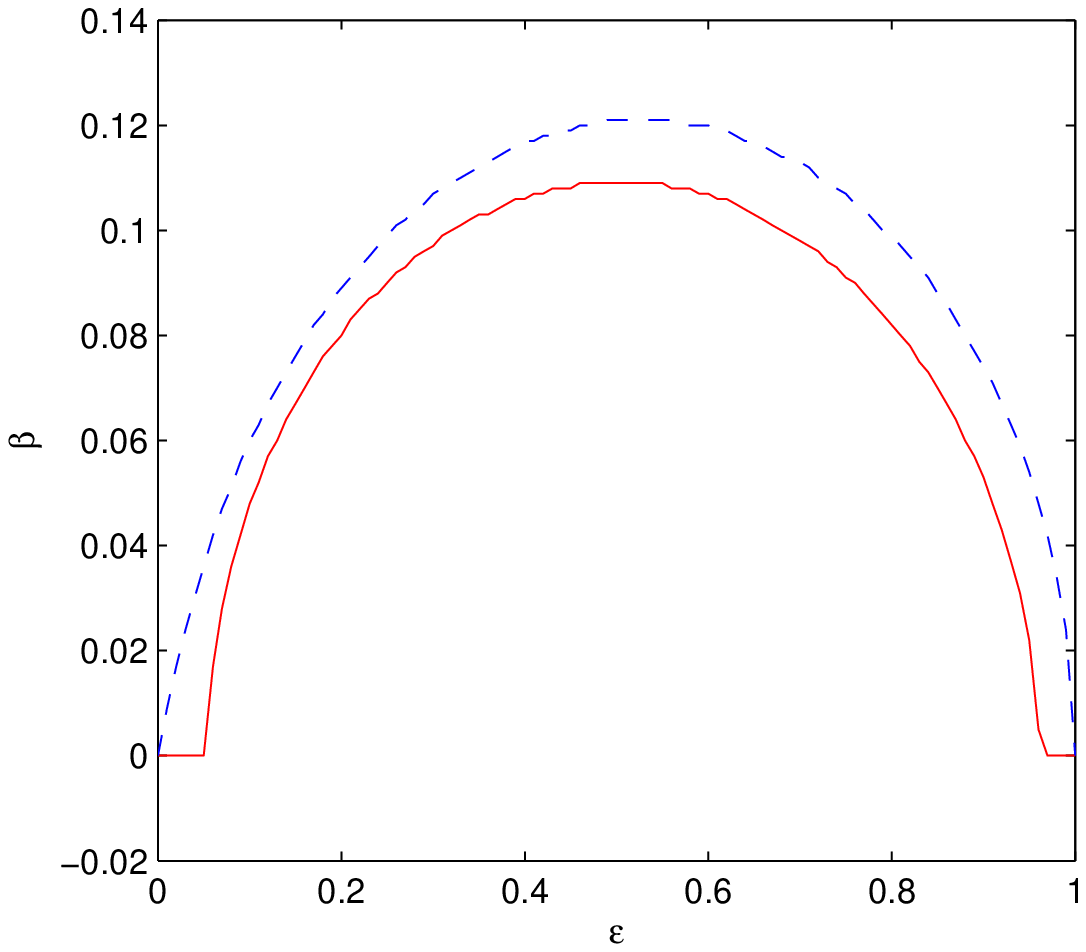}}
\subfigure[``Two-dimensional'' parameter regime.] {
\includegraphics[scale=0.5]{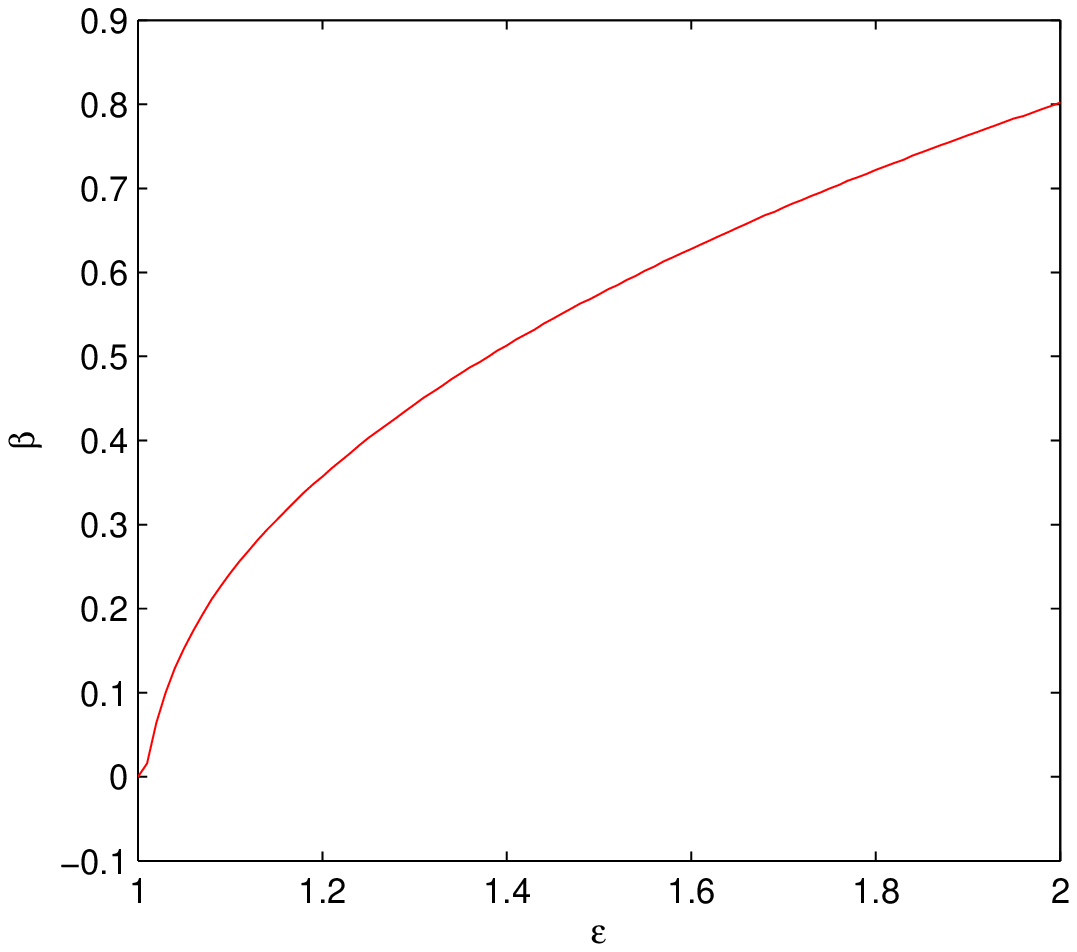}}
\caption{The biffurcation diagram $\beta$ vs. $\epsilon$. The dashed
line indicates the appearance of one real negative eigenvalue of the
matrix $J_N$, which happens only in the regime $0 < \epsilon < 1$
(a). The solid line shows the point at which the real part of two
conjugate complex eigenvalues of $J_N$ become negative. This
bifurcation disappears in the ``three-dimensional'' parameter regime
(a) at $0 \le \epsilon \le 0.05$ and $0.97 \le \epsilon \le 1$.
Observe that for positive viscosity $\nu$ the solution $\B u^N$ is
linearly stable for every $N$ at $\epsilon = 1$, which can be shown
rigorously.}
\end{center}
\end{figure}

For other values of the parameters, we substitute $\lambda = 2$, and
write the characteristic polynomial of the matrix $J_N$
\begin{align}
x^4 - \frac{325}{16} \beta x^3 & + \Big (4 \epsilon^2 - \frac{19}{4}
\epsilon + 1 + \frac{4497}{64} \beta^2 \Big ) x^2 - \notag
\\
& - \Big (\frac{325}{16} \beta^2 + \frac{5}{4} \epsilon^2 +
\frac{257}{16} - \frac{197}{16} \epsilon \Big ) \beta x + \notag
\\
& + (\epsilon^3 - \epsilon^2) + \Big ( 1 + \frac{239}{16} \epsilon +
\frac{1}{16} \epsilon^2 + \beta^2 \Big ) \beta^2 = 0.
\label{charpoly}
\end{align}
Next, by fixing an $\epsilon$, we find the largest $\beta$ when the
real part of the roots of the polynomial (\ref{charpoly}) changes
its sign. The result of this calculation is shown at Figure~$1$.

For every $\epsilon$ in the ``three-dimensional'' subrange of
parameters $0.05 < \epsilon < 0.97$, there are two bifurcation
points. First, there exists a value of $\beta$ for which one of the
real eigenvalues of $J_N$ crosses $0$ and becomes negative. Further
decreasing $\beta$ we observe another bifurcation at which real part
of a pair of complex conjugate eigenvalues becomes negative.
Therefore, for $0 < \epsilon < 1$ there exists a function
$m(\epsilon)$, such that the matrix $J_N$ has $3$ negative
eigenvalues for
\begin{equation}
0 < \beta = \nu k_N^4 \le m(\epsilon),
\end{equation}
or in other words, for $N$ satisfying
\begin{equation} \label{eq_n_bound3}
4 N \le \log_\lambda \nu^{-1} + \log_\lambda m(\epsilon),
\end{equation}
Note, that $m(\epsilon) > 0$, at $\epsilon \in (0.05, 0.97)$, and
$m(\epsilon) = 0$ otherwise.

For the range of parameters $0 < \epsilon \le 0.05$ and $0.97 \le
\epsilon < 1$ we observe only one bifurcation point at which one of
the real eigenvalues becomes negative. Therefore, we call this
regime -- the ``non-chaotic'' range of parameters and the reason for
that will be explained further.

Finally, for the ``two-dimensional'' range of parameters $1 <
\epsilon < 2$ the scenario is a little different, as we observe only
one bifurcation point at which real part of a pair of complex
conjugate eigenvalues becomes negative. Namely, for $1 < \epsilon <
2$ there exists a function $d(\epsilon)$, such that the matrix $J_N$
has $2$ negative eigenvalues for
\begin{equation}
0 < \beta = \nu k_N^4 \le d(\epsilon),
\end{equation}
or equivalently, for $N$ satisfying
\begin{equation} \label{eq_n_bound2}
4 N \le \log_\lambda \nu^{-1} + \log_\lambda e(\epsilon).
\end{equation}
In this case we also have, $e(\epsilon) > 0$ for all $\epsilon \in
(1, 2]$, and $d(1) = 0$.

\subsection{Calculating the lower bound of the dimension of the attractor}
\label{sec:lower}

In the previous section we show that the Sabra shell model has at
most three unstable direction for the ``single-mode'' forcing.
Therefore, we need other type of the force to get a large number of
unstable direction, and finally, the lower-bound for the dimension
of the global attractor, which would be close to the upper-bound
calculated in~\cite{CLT05}. Let us define the forcing
\begin{equation} \label{gforce}
\B g = \sum_{j=1}^{\infty} \B f^{5j}, \;\;\; \abs{\B g} = \nu
\frac{1}{\lambda^{5} \sqrt{1-\lambda^{-10}}},
\end{equation}
where $\B f^{5j}$ is defined in (\ref{eq_single_mode_forcing}). Then
the stationary solution corresponding to that forcing is
\begin{equation}
\B u^{\B g} = \sum_{j=1}^{\infty} \B u^{5j},
\end{equation}
where $\B u^{5j}$ is defined in (\ref{eq_kolmogorov_flow}). Using
the results of the previous section on the stability of the
single-mode stationary solution we conclude that for $0 < \epsilon <
1$, the number of the unstable directions of the solution $\B u^{\B
g}$ equals to $3 N /5$, where $N$ satisfies the  relation
(\ref{eq_n_bound3}). On the other hand, the number of the unstable
directions of the solution $\B u^{\B g}$ for $1 < \epsilon < 2$
equals to $2 N /5$, where $N$ satisfies relation
(\ref{eq_n_bound2}).

Recall the definition of the generalized Grashoff number
(\ref{eq_grashof}), which  in our case satisfies
\begin{align*}
G = \frac{\abs{\B g}}{\nu^2 \lambda^3} = \frac{1}{\nu \lambda^{8}
\sqrt{1-\lambda^{-10}}}.
\end{align*}
Therefore, we can rewrite the bounds (\ref{eq_n_bound3}) and
(\ref{eq_n_bound3}) in terms of the generalized Grashoff number to
obtain
\begin{equation}
4 N = \log_\lambda \nu^{-1} + \log_\lambda f(\epsilon) \le
\log_\lambda G + \log_\lambda f(\epsilon),
\end{equation}
where $f(\epsilon)$ denotes $m(\epsilon)$ or $d(\epsilon)$.
Therefore, we proved the following statement.

\begin{theorem}
The Hausdorff and fractal dimensions   of the global attractor
$\mathcal A$ of the Sabra shell model of turbulence with $\nu > 0$
and the forcing defined in (\ref{gforce}) satisfies
\begin{equation}
\dim_F(\mathcal{A}) \ge \dim_H(\mathcal{A}) \ge K \log_\lambda G +
\log_\lambda f(\epsilon),
\end{equation}
for the positive constant $K$ depending on $\epsilon$ satisfying
\begin{equation}
K(\epsilon) = \left\{
  \begin{array}{ll}
    \frac{3}{20}, & \text{for } \;\; 0.05 < \epsilon < 0.97, \\
    \frac{1}{10}, & \text{for } \;\; 1 < \epsilon \le 2, \\
    \frac{1}{20}, & \text{for } \;\; 0 < \epsilon \le 0.05, \;\;\; \text{and} \;\; 0.97 \le \epsilon < 1, \\
  \end{array}
\right.
\end{equation}
and some positive real function $f(\epsilon)$, which is $0$ only for
$\epsilon = 1$.

\end{theorem}

The lower bounds for the global attractor, given by the last Theorem
do not match exactly the upper bounds which were obtained previously
in \cite{CLT05}, namely
\begin{equation} \label{upper}
\dim_H(\mathcal{A}) \le \dim_F(\mathcal{A}) \le \frac{1}{2}
\log_\lambda G - C(\epsilon),
\end{equation}
where the function $C(\epsilon)$ stays positive and bounded for
every $\epsilon\in (0, 2)$. Moreover, the constant $K$ in front of
the $\log_\lambda G$ term, although can be slightly improved, cannot
be brought much closer to $\half$ to match the upper bound of
(\ref{upper}).

\section{Existence of a trivial global attractor for any value of $\nu$}
\label{sec:example}

It is well known that the attractor for the $2$-dimensional
space-periodic Navier-Stokes equation with a particular form of the
forcing can consists of only one function. This well-known example
is due to Yudovich \cite{Yu65_Example} and independently by
Marchioro \cite{Ma86_TrivialAttr} (for the proof see also
\cite{FMRT01}). The same is true for the Sabra shell model for $1 <
\epsilon < 2$, therefore, we need to stress that the bounds that we
obtained for the dimension of the global attractor are valid only
for the particular type of forcing that we used in our calculations.

We mentioned in the introduction that for the $2$-dimensional
parameters regime the inviscid Sabra shell model without forcing
conserves the following quantity
\[
\abs{A^{\gamma} u}^2 = \sum_{n=1}^\infty k_n^{4\gamma} \abs{u_n}^2,
\]
for $4\gamma = - \log_\lambda (\epsilon - 1)$. For $m > 0$ we denote
by $\B P_m$ -- the projection onto the first $m$ coordinates of the
sequence $\B u$, and $\B Q_m = \B I - \B P_m$.

\begin{theorem} \label{thm_dm_forcing2}
Suppose that the forcing $\B f$ acts only on the $N$-th shell for
some $N \ge 1$. Let $\B u(t)$ be the solution of the the equation
(\ref{eq_abstract_model}) in the ``two-dimensional'' regime of
parameters $1 < \epsilon < 2$, such that for $\gamma = - \frac{1}{4}
\log_\lambda (\epsilon - 1)$ we have
\begin{equation*}
Re (\B B(\B u, \B u), \B A^{2 \gamma}\B u) = 0.
\end{equation*}
Then
\begin{equation}\label{eq_dm_lemma2}
\limsup_{t\to \infty} \abs{\B Q_m \B u(t)}^2 \le C
\frac{1}{k_{m+1}^{4\gamma}} \liminf_{t\to \infty} \abs{\B P_m \B
u(t)}^2,
\end{equation}
for $C = \frac{k_N^{4\gamma} - k_1^{4\gamma}}{1 -
\lambda^{-4\gamma}}$ and $m \ge N$.
\end{theorem}

\begin{proof}
Taking the scalar product of the equation (\ref{eq_abstract_model})
with $\B u$ and with $\B A^{2 \gamma}\B u$ we get two equations
\begin{equation*} \label{eq_dm_first_eq2}
\half \ddt \abs{\B u}^2 + \nu (\B A\B u, \B u) = Re(f_N u_N^*),
\end{equation*}
and
\begin{equation*} \label{eq_dm_second_eq2}
\half \ddt \abs{\B A^{\gamma} \B u}^2 + \nu (\B A \B u, \B A^{2
\gamma} \B u) = Re(k_N^{4\gamma} f_N u_N^*).
\end{equation*}
Multiplying the energy equation by $k_N^{4\gamma}$ and subtracting
it from the last equation we get
\begin{equation} \label{eq_est2}
\half \ddt (\abs{\B A^{\gamma} \B u}^2 - k_N^{4\gamma} \abs{\B u}^2)
+ \nu (\abs{\B A^{\gamma + 1/2} \B u}^2 - k_N^{4\gamma} \nm{\B u}^2)
= 0.
\end{equation}
On the other hand,
\begin{align*}
\abs{\B A^{\gamma + 1/2} \B u}^2 - k_N^{4\gamma} \nm{\B u}^2 & =
\sum_{n=1}^\infty k_n^{2} (k_n^{4\gamma} - k_N^{4\gamma})
\abs{u_n}^2 \ge
\\
& \ge k_N^{2} \sum_{n=1}^\infty (k_n^{4\gamma} - k_N^{4\gamma} )
\abs{u_n}^2 = k_N^{2} (\abs{\B A^{\gamma} \B u}^2 - k_N^{4\gamma}
\abs{\B u}^2).
\end{align*}
Plugging the last expression into (\ref{eq_est2}) yields
\[
\ddt (\abs{\B A^{\gamma} \B u}^2 - k_N^{4\gamma} \abs{\B u}^2) \le
-2 \nu k_N^2 (\abs{\B A^{\gamma} \B u}^2 - k_N^{4\gamma} \abs{\B
u}^2),
\]
and therefore,
\begin{equation}\label{eq_dm_third2}
\limsup_{t\to \infty} (\abs{\B A^{\gamma} \B u(t)}^2 - k_N^{4\gamma}
\abs{\B u(t)}^2) = 0.
\end{equation}

Finally,
\begin{align*}
\abs{\B Q_m \B u}^2 & = \sum_{i=m+1}^\infty \abs{u_i}^2 =
\sum_{i=m+1}^\infty \frac{k_i^{4\gamma} -
k_N^{4\gamma}}{k_i^{4\gamma} - k_N^{4\gamma}} \abs{u_i}^2 \le
\\
& \le \frac{1}{k_{m+1}^{4\gamma} - k_N^{4\gamma}} (\abs{\B Q_m \B
A^{\gamma} \B u}^2 - k_N^{4\gamma} \abs{\B Q_m \B u}^2) =
\\
& = \frac{1}{k_{m+1}^{4\gamma} - k_N^{4\gamma}} \bigg( (\abs{\B
A^{\gamma} \B u}^2 - k_N^{4\gamma} \abs{\B u}^2) - (\abs{\B P_m \B
A^{\gamma} \B u}^2 - k_N^{4\gamma} \abs{\B P_m \B u}^2) \bigg) \le
\\
& \le \frac{1}{k_{m+1}^{4\gamma} - k_N^{4\gamma}} \bigg( (\abs{\B
A^{\gamma} \B u}^2 - k_N^{4\gamma} \abs{\B u}^2) + (k_N^{4\gamma} -
k_1^{4\gamma}) \abs{\B P_m \B u}^2 \bigg),
\end{align*}
and the result follows from (\ref{eq_dm_third2}).

\end{proof}

\begin{corollary}
The global attractor of the Sabra shell model of turbulence in the
``two-dimensional'' regime of parameters $1 < \epsilon < 2$ with the
force applied only to the first shell
\begin{equation} \label{1force}
\B f^1 = (f, 0, 0, \dots),
\end{equation}
is reduced to a single stationary solution
\begin{equation*}
\B u^1 = (\frac{f}{\nu k_1^2}, 0, 0, \dots).
\end{equation*}
\end{corollary}

\begin{proof}
Let $\B u = (u_1, u_2, \dots)$ be a solution of the Sabra shell
model with the forcing $\B f$ defined by (\ref{1force}). Then it
immediately follows from Theorem~\ref{thm_dm_forcing2} that
\begin{equation*}
\limsup_{t\to \infty} \abs{\B Q_1 \B u}^2 = 0,
\end{equation*}
which means that $\limsup_{t\to \infty} \abs{u_n} = 0$, for every
$n\ge 2$.

Define $\B v = (v_1, v_2, \dots)$ as  $\B v = \B u - \B u^1$, which
satisfies the equation
\begin{equation*}
\pdt{\B v} + \nu \B A \B v + \B B(\B u, \B u) = 0,
\end{equation*}
where we used the fact that $\B B(\B u^1, \B u^1) = 0$. Taking the
inner product of the equation with the vector $\B P_1 \B v = (v_1,
0, 0, \dots)$ we get that $\abs{v_1(t)}^2$ satisfies
\begin{equation*}
\half \ddt \abs{v_1(t)}^2 + \nu k_1^2 \abs{v_1(t)}^2 + v_1(t) u_2(t)
u_3(t) = 0.
\end{equation*}
Using the fact that $u_2(t), u_3(t)$ tend to $0$ as $t\to \infty$ we
conclude that $\abs{v_1(t)}^2\to 0$ as $t\to \infty$. Therefore,
\begin{equation*}
\limsup_{t\to \infty} \abs{\B v}^2 = \limsup_{t\to \infty} \abs{\B u
- \B u^1}^2 = 0.
\end{equation*}
finishing the proof.
\end{proof}

\section{Conclusion}

In this work we obtained lower bounds for the dimension of the
global attractor of the Sabra shell model of turbulence for specific
choices of the forcing term. Our main result states that for these
specific choices of the forcing term the Sabra shell model has a
large attractor for all values of the governing parameter
$\epsilon\in (0, 2) \setminus \{1\}$. We also showed the scenario of
the transition to chaos in the model, which is slightly different
for the two- and three-dimensional parameters regime. In addition,
in the three-dimensional parameters regime, $\epsilon\in (0, 1)$, we
found that when the parameter $\epsilon$ becomes sufficiently close
to $0$ or to $1$ where the chaotic behavior in the vicinity of the
stationary solution changes dramatically.

Finally, we show that in the ``two-dimensional'' parameters regime
the Sabra shell model has a trivial attractor reduced to a single
equilibrium solution for any value of viscosity $\nu$, when the
forcing is applied only to the first shell. This result is true also
for the two-dimensional NSE due to Yudovich \cite{Yu65_Example} and
independently by Marchioro \cite{Ma86_TrivialAttr}  (see also
\cite{FMRT01}).

\section*{Acknowledgments}

The work of P.C. was supported in part by the NSF grant
No.~DMS--0504213. The work of E.S.T. was also supported in part by
the NSF grant No.~DMS--0504619, the ISF grant No.~120/06, and by the
BSF grant No.~2004271.


\begin{thebibliography}{99}


\bibitem{BV83_AttractorDim}
  A. V. Babin, M. I. Vishik,
  \textit{Attractors of partial differential equations and estimates of their dimension},
  Uspekhi Mat. Nauk, \textbf{38} (1983), 133-187 (in Russian); Russian Math. Surveys, \textbf{38}, 151-213 (in English).


\bibitem{Bi03}
  L. Biferale,
  \textit{Shell models of energy cascade in turbulence},
  Annual Rev. Fluid Mech., \textbf{35} (2003), 441-468.


\bibitem{BiLLP95}
  L. Biferale, A. Lambert, R. Lima, G. Paladin,
  \textit{Transition to chaos in a shell model of turbulence},
  Physica D, \textbf{80} (1995), 105--119.


\bibitem{BJPV98}
  T. Bohr, M. H. Jensen, G. Paladin, A. Vulpiani,
  \textit{Dynamical Systems Approach to Turbulence},
  Cambridge University press, 1998.


\bibitem{CLT05}
  P. Constantin, B. Levant, E. S. Titi,
  \textit{Analytic study of the shell model of turbulence},
  Physica D, \textbf{219} (2006), 120-141.


\bibitem{CLT05_Inviscid}
  P. Constantin, B. Levant, E. S. Titi,
  \textit{A note on the regularity of inviscid shell models of turbulence},
  submitted.


\bibitem{FMRT01}
  C. Foias, O. Manley, R. Rosa, R. Temam,
  \textit{Navier-Stokes Equations and Turbulence},
  Cambridge University press, 2001.


\bibitem{Fr95}
  U. Frisch,
  \textit{Turbulence: The Legacy of A. N. Kolmogorov},
  Cambridge University press, 1995.


\bibitem{Gl73}
  E. B. Gledzer,
  \textit{System of hydrodynamic type admitting two quadratic
           integrals of motion},
  Sov. Phys. Dokl., \textbf{18} (1973), 216-217.


\bibitem{KaLoSch97}
  L. Kadanoff, D. Lohse, N. Schr\"{o}ghofer,
  \textit{Scaling and linear response in the GOY turbulence model},
  Physica D, \textbf{100} (1997), 165--186.


\bibitem{KoOkJe98}
  J. Kockelkoren, F. Okkels, M. H. Jensen,
  \textit{Chaotic behavior in shell models and shell maps},
  J. Stat. Phys., \textbf{93}, (1998), 833.


\bibitem{LL77_Hydrodynamics}
  L. D. Landau, E. M. Lifschitz,
  \textit{Fluid Mechanics},
  Pergamon, Oxford 1977.


\bibitem{Liu93_Attractor}
  V. X. Liu,
  \textit{A sharp lower bound for the Hausdorff dimension of the global attractors of the 2D Navier-Stokes equations},
  Commun. Math. Phys., \textbf{158}, (1993), 327-339.


\bibitem{LP98}
  V. S. L'vov, E. Podivilov, A. Pomyalov, I. Procaccia, D. Vandembroucq,
  \textit{Improved shell model of turbulence},
  Physical Review E., \textbf{58 (2)} (1998), 1811-1822.


\bibitem{Ma86_TrivialAttr}
  C. Marchioro,
  \textit{An example of absence of turbulence for any Reynolds number},
  Comm. Math. Phys., \textbf{105} (1986), 99-106.


\bibitem{MeSi61_Kolmogorov}
  L. D. Meshalkin and Y. G. Sinai,
  \textit{Investigation of the stability of a stationary solution of a system of equations for the plane movement of an incompressible viscous liquid},
  J. Appl. Math. Mech., \textbf{25} (1961), 1700--1705.


\bibitem{OY89}
  K. Okhitani, M. Yamada,
  \textit{Temporal intermittency in the energy cascade process and
           local Lyapunov analysis in fully developed model of turbulence},
  Prog. Theor. Phys., \textbf{89} (1989), 329-341.


\bibitem{Te88}
  R. Temam,
  \textit{Infinite-Dimensional Dynamical Systems in Mechanics and Physics},
  Springer-Verlag, New-York, 1988.


\bibitem{YO87_3D}
  M. Yamada, K. Okhitani,
  \textit{Lyapunov spectrum of a chaotic model of three-dimensional turbulence},
  J. Phys. Soc. Jpn., \textbf{56} (1987), 4210--4213.


\bibitem{YO88_2D}
  M. Yamada, K. Okhitani,
  \textit{Lyapunov spectrum of a model of two-dimensional turbulence},
  Phys. Rev. Let., \textbf{60 (11)} (1988), 983--986.


\bibitem{YO98_Formulas}
  M. Yamada, K. Okhitani,
  \textit{Asymptotic formulas for the Lyapunov spectrum of fully developed shell model turbulence},
  Phys. Rev. E, \textbf{57 (6)} (1998), 57--60.


\bibitem{Yu65_Example}
  V. I. Yudovich,
  \textit{Example of the generation of a secondary stationary or periodic flow when there is
             loss of stability of the laminar flow of a viscous incompressible fluid},
  J. Appl. Math. Mech., \textbf{29} (1965), 527–-544.


\end{thebibliography}
\end{document}